# Further evidence for energy sorting from the even-odd effect in fission-fragment element distributions

Karl-Heinz Schmidt, Beatriz Jurado

*CENBG, CNRS/IN2P3, Chemin du Solarium B. P. 120, 33175 Gradignan, France*

**Abstract.** The even-odd effect in fission is explained by a model based on statistical mechanics. It reveals that the variation of the even-odd effect with the mass of the fissioning nucleus and the increase towards asymmetric splits is due to the important statistical weight of configurations where the light fission fragment populates the ground state of an even-even nucleus. This implies that entropy drives excitation energy and unpaired nucleons predominantly to the heavy fragment. Therefore, the even-odd effect is an additional signature of the recently discovered energy-sorting process in fission.

Nuclear fission is a large-scale collective motion where a heavy nucleus gradually evolves into two nuclei. Due to the strong mutual Coulomb repulsion, the emerging fission fragments separate at scission and move apart with high kinetic energies. The fully accelerated fission fragments de-excite by neutron and gamma emission. Low-energy fission offers a unique possibility to study the behavior of two moderately excited quantum-mechanical systems in contact under the influence of strong pairing correlations. Recent experiments show that nuclear level densities at low excitation energy follow rather well the constant-temperature behavior where the inverse logarithmic slope of the level density essentially remains constant with excitation energy [1, 2]. This behavior results from the combined influence of pairing correlations and the continuous melting of Cooper pairs [2]. We have shown that the constant-temperature regime of the level density leads to a surprising effect where practically all the excitation energy that is available before scission is transferred to the heavy fragment [3]. This process of energy sorting explains why an increase of the available excitation energy leads to an increase of the number of prompt neutrons emitted by the heavy fragment, only. We have investigated the energy sorting on the basis of statistical mechanics in [4, 5]. In this work, we further use statistical mechanics to show that the asymmetry dependence of the even-odd effect in fission is an additional signature of the mechanism of energy sorting.

Fig, 1 shows the measured fission-fragment yields as a function of proton number ($Z$) for the fissioning nucleus $^{229}$Th, which was excited in the Coulomb field of lead target atoms slightly above the fission barrier [6]. The global shape of the data from Fig. 1 can be described by three humps, one centered at symmetry ($Z\approx45$) and two at asymmetry ($Z\approx36$ and 54). These humps result from the shape of the potential as a function of mass (or charge) asymmetry as given by the liquid-drop model with influence of shell effects [7]. The even-odd effect in fission-fragment yields is the fine structure showing an enhanced production of fragments with even $Z$ that is superimposed to the gross shape of the yields. It was proposed in [8] to quantify the even-odd staggering by the local even-odd effect $\delta_p(Z)$, which corresponds to the third differences of the logarithm of the yields $Y$:

$$\delta_p(Z) = \frac{1}{8}(-1)^Z[\ln Y(Z+3) - \ln Y(Z) - 3(\ln Y(Z+2) - \ln Y(Z+1))] \quad (1)$$

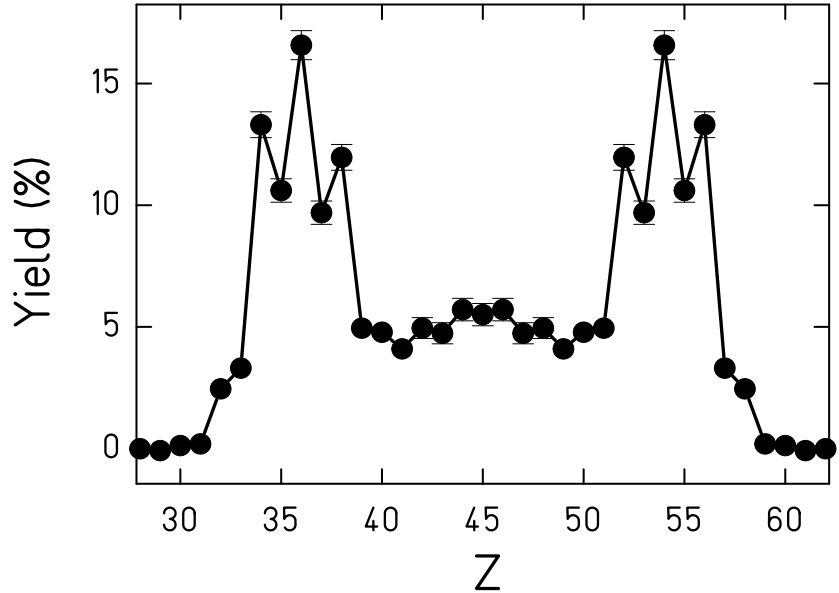

**Figure 1.** Element distribution observed in the electromagnetic-induced fission of $^{229}$Th [6].

$\delta_p(Z)$ filters out from the yields the variations that extend over a significant number of charges and are caused by the global shape of the potential energy. We can consider two curves; one links the logarithms of the yields of neighboring even-$Z$ fragments $\ln Y_{even}(Z)$, and the other connects the logarithms of the yields of neighboring odd-$Z$ fragments $\ln Y_{odd}(Z)$. These two curves are continuous functions (if the yields $Y_{even}$ and $Y_{odd}$ follow a Gaussian shape, the curves are parabolas) and can be evaluated for any value of $Z$. As explained in [8], $\delta_p(Z)$ equals half the distance between the two curves:

$$\delta_p(Z) = 0.5(\ln Y_{even}(Z) - \ln Y_{odd}(Z)) \quad (2)$$

Fig. 2 shows experimental results on $\delta_p(Z)$ as a function of charge asymmetry for different fissioning nuclei. In these data fission was induced by thermal neutrons. Fig. 2 illustrates several general trends:

(i) The amplitude of $\delta_p$ decreases with increasing mass of the fissioning system.
(ii) For a given fissioning nucleus $\delta_p$ increases with asymmetry.
(iii) Also odd-$Z$ fissioning systems like $^{243}$Am show an even-odd effect at large asymmetry whose magnitude is about the same as for even-$Z$ systems of comparable mass.

As described in [9], presently there is no consistent explanation for all these features. In the following, we present a simple model for the even-odd effect that is based on statistical mechanics.

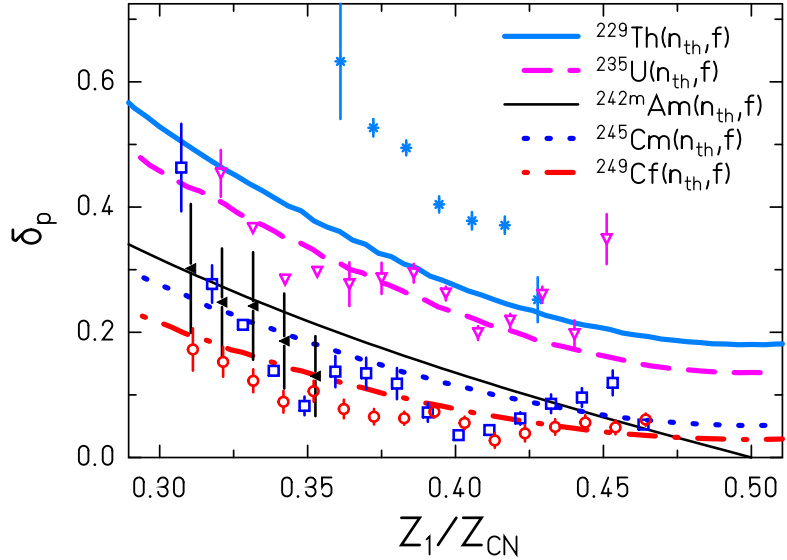

**Figure 2:** (Color online) Local even-odd effect $\delta_p$ as a function of asymmetry, parameterized as the ratio of the charge of the light fragment $Z_1$ and the charge of the fissioning nucleus $Z_{CN}$. The symbols represent the experimental data which are taken from the compilation shown in [9]. The lines correspond to the results of the model developed in this work.

Because of the minimum number of microstates available above the outer fission barrier, the excited nucleus can only pass the barrier with a sizeable probability if the different types of degrees of freedom (intrinsic and collective) are in statistical equilibrium [10]. This implies that the available energy above the fission barrier, which is on the average equal to the total energy of the nucleus minus the height of the outer fission barrier, is equally shared between the different degrees of freedom. Most of the available states are intrinsic excitations. Thus, most of the excitation energy available above the outer saddle is stored in intrinsic excitations. The intrinsic excitation energy grows on the way from saddle to scission because part of the potential-energy release is dissipated into intrinsic excitations. Two-centre shell-model calculations documented in Fig. 12 of [11] show that there are many level crossings on the first section behind the outer saddle. Afterwards, the single-particle levels change only little. Level crossings lead to intrinsic excitations [12]. Consequently, the additional excitation energy is mostly dissipated in the vicinity of the outer saddle. The dissipated energy increases with the mass of the fissioning nucleus because the fission barrier is located at smaller deformations, and the range with a high number of level crossings is extended. Additional intrinsic excitation may appear at neck rupture. Theoretical investigations of the gradual transition from the mononucleus regime to the dinuclear system concerning shell effects [11], pairing correlations [13] and congruence energy [14] show that the properties of the individual fission fragments are already very well established in the vicinity of the outer saddle. Therefore, close to the outer saddle the fissioning system consists of two well-defined nuclei in contact through the neck and a total amount of excitation energy $E_{tot}$ that is equal to the intrinsic excitation energy above the outer saddle plus the energy acquired by dissipation on the first section behind the outer saddle. Intrinsic excitations are expected to be homogeneously distributed within the nuclear volume. This is likely to hold also in the transition from a mononuclear to a dinuclear system that takes place very rapidly near the outer saddle [11]. Consequently, a reasonable assumption is that $E_{tot}$ is initially shared among the fragments according to the ratio of their masses.

We assume that the system formed by the two nuclei in contact then evolves to a state of statistical equilibrium, the macrostate of maximum entropy, where all the available microstates have equal probability [15]. This implies that the intrinsic excitation energy will be distributed among the two nascent fragments according to the probability distribution of the available microstates which is given by the total nuclear level density[1]. Energy sorting will take place, and the light fragment will transfer essentially all its excitation energy to the heavy one [3, 5]. Nucleon exchange between the fragments establishes also an equilibrium between even-even, even-odd, odd-even, and odd-odd light and heavy nascent fragments with a restriction on the gross mass asymmetry given by the bottom of the potential in the fission valleys and on the total intrinsic excitation energy $E_{tot}$. To obtain the probability of populating a given configuration at statistical equilibrium and derive the local even-odd effect $\delta_p$ we have to consider the level densities of neighboring even-even, even-odd, odd-even or odd-odd nuclei in an absolute energy scale. However, as said above, the quantity $\delta_p$ (eq. 1) filters out the slowly-varying components of the yields. Therefore, we have to use level densities where these effects are filtered out as well. As illustrated when discussing the meaning of $\delta_p$, we can consider a smooth surface (in the neutron number $N$ and $Z$ space) that connects the yields of odd-odd fragments. This surface can be associated with the potential energy in the fission valley corresponding to odd-odd pre-fragments. The potential surface of odd-mass nuclei is at an energy $-\Delta$ ($\Delta$ being the pairing gap $\approx 12/\sqrt{A}$, $A$ is the mass number) with respect to the surface of odd-odd nuclei, and the potential surface of even-even nuclei is at an energy of $-2\Delta$. Thus, the required filtering of the level densities can be obtained by placing the level densities in an excitation-energy scale that is reduced by $2\Delta$ for even-even nuclei, by $\Delta$ for even-odd and odd-even nuclei and is left unchanged for odd-odd nuclei: $E_{gs}$-n$\Delta$. We have done this on Fig. 3 with experimental level densities determined by the Oslo method [16] of various even-even, odd-A and odd-odd nuclei located in two distinct mass regions around $A$=165 and 45. These two groups of nuclei are representative of the complementary fission fragments produced in very asymmetric fission. Within the heavy group, the level densities of neighboring even-even and even-odd nuclei are almost identical. Sizeable differences appear only in the energy interval -2$\Delta_2$<$E$<-$\Delta_2$, where only even-even nuclei have states. For the light-mass group, the level densities converge well at positive energies. Significant fluctuations are present at negative reduced energies due to structure effects since these nuclei are relatively light and located between the $Z$=20 and $Z$=28 shell closures. Fig. 3 clearly shows that the logarithmic slope of the level densities is nearly constant and that the logarithmic slope of the heavy group is much larger than the one of the light group.

---

[1] The degeneracy of magnetic substates is not considered, because it contributes very little to the variation of the density of states as a function of excitation energy.

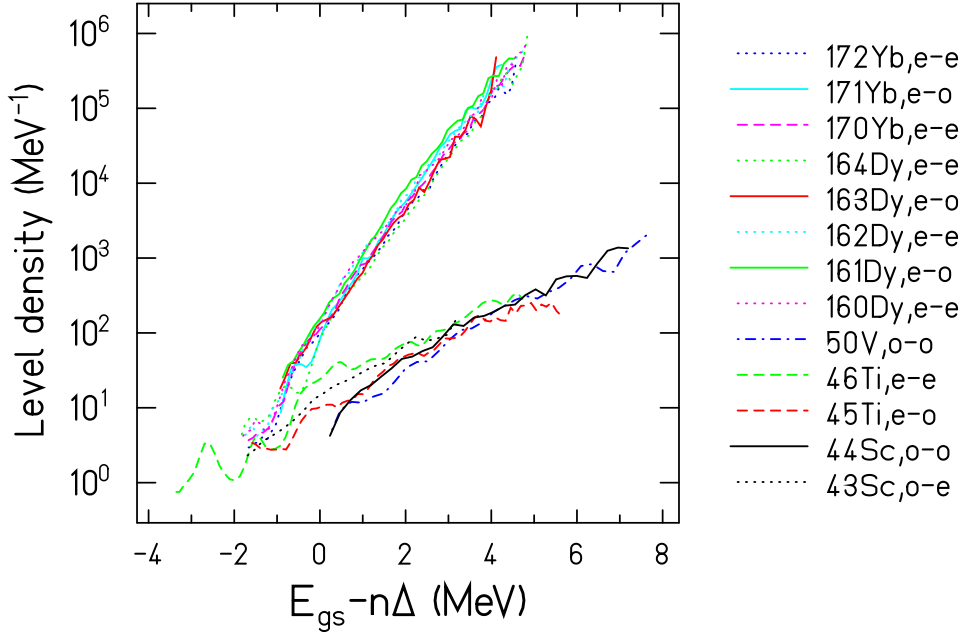

**Figure 3.** (Color online) Experimental level densities of various nuclei [1, 17-23]. The excitation energy is reduced by 2Δ for even-even (e-e) nuclei, by Δ for even-odd (e-o) or odd-even (o-e) nuclei and left unchanged for odd-odd (o-o) nuclei.

The total amount of possible configurations (or microstates) with particular values of $Z_1$ and $Z_2$ is directly related to the integral of the total level density for that particular split over the excitation energy of one fragment $E_i$, with the condition that $E_1+E_2=E_{tot}$. This integral reflects the freedom of the system in the division of excitation energy (note that $E_1$ and $E_2$ are defined in the reduced energy scale, relative to the potential surface of odd-odd nuclei). The total level density is given by the product of the level densities of the two fragments $\rho_1(E_1)\rho_2(E_2)$. The most probable configuration is the one that provides the highest total level density. Starting from a situation where $E_1$ and $E_2$ are in the range of energy shown in Fig. 3, due to the very different logarithmic slopes of the level densities for the heavy and the light fragment group, the most favorable configurations are those that minimize the excitation energy of the light fragment. In other words, energy sorting takes place [3]. Only at the end of the energy sorting, when $E_1 \approx 0$ and $E_2 \approx E_{tot}$, the benefit of transferring the unpaired nucleons to the heavy fragment to form an even-odd, an odd-even or an even-even light fragment becomes apparent because only for these configurations there are states available. If an even-even light fragment in the ground state is formed, instead of an odd-odd one, the energy in the heavy fragment increases to $E_2=E_{tot}+2\Delta_1$, which for the nuclei considered in Fig. 3 corresponds to an increase of the level density of the heavy fragment of more than three orders of magnitude, while $E_1=-2\Delta_1$. It becomes clear that configurations with fully paired light fragments (light fragments with no quasi particle excitations) are strongly favored for very asymmetric fission. Because the logarithmic slope of the constant-temperature level densities is roughly proportional to $A^{2/3}$ [24], for less asymmetric splits the logarithmic slopes of the level densities of the two fragments will become closer. Therefore, the probability to populate the ground states of even-even or even-odd light fragments decreases with decreasing asymmetry.

For an even-even fissioning nucleus, the number of configurations with $Z_1$ even is given by:

$$N^{ee}_{Z_1=e}(Z_1) = \int_{-2\Delta_1}^{E_{tot}+2\Delta_2} \rho_1(E_1)_{(ee)} \rho_2(E_{tot} - E_1)_{(ee)} dE_1 + \int_{-\Delta_1}^{E_{tot}+\Delta_2} \rho_1(E_1)_{(eo)} \rho_2(E_{tot} - E_1)_{(eo)} dE_1 \quad (3)$$

where $\rho_i(E_i)_{(ee)}$ and $\rho_i(E_i)_{(eo)}$ are the level densities of representative even-even and even-odd nuclei, respectively, with mass close to $A_1$ or $A_2$. The number of configurations with $Z_1$ odd is:

$$N^{ee}_{Z_1=o}(Z_1) = \int_{-\Delta_1}^{E_{tot}+\Delta_2} \rho_1(E_1)_{(oe)} \rho_2(E_{tot} - E_1)_{(oe)} dE_1 + \int_{0}^{E_{tot}} \rho_1(E_1)_{(oo)} \rho_2(E_{tot} - E_1)_{(oo)} dE_1 \quad (4)$$

where $\rho_i(E_i)_{(oe)}$ and $\rho_i(E_i)_{(oo)}$ are the level densities of representative odd-even and odd-odd nuclei, respectively, with mass close to $A_1$ or $A_2$. The yield for even-$Z_1$ nuclei is $Y^{ee}_{Z_1=e}(Z_1) = N^{ee}_{Z_1=e}(Z_1) / N^{ee}_{tot}(Z_1)$ with $N^{ee}_{tot}(Z_1) = N^{ee}_{Z_1=e}(Z_1) + N^{ee}_{Z_1=o}(Z_1)$. For an odd-even fissioning nucleus, we have:

$$N^{oe}_{Z_1=e}(Z_1) = \int_{-2\Delta_1}^{E_{tot}+\Delta_2} \rho_1(E_1)_{(ee)} \rho_2(E_{tot} - E_1)_{(oe)} dE_1 + \int_{-\Delta_1}^{E_{tot}} \rho_1(E_1)_{(eo)} \rho_2(E_{tot} - E_1)_{(oo)} dE_1 \quad (5)$$

$$N^{oe}_{Z_1=o}(Z_1) = \int_{-\Delta_1}^{E_{tot}+2\Delta_2} \rho_1(E_1)_{(oe)} \rho_2(E_{tot} - E_1)_{(ee)} dE_1 + \int_{0}^{E_{tot}+\Delta_2} \rho_1(E_1)_{(oo)} \rho_2(E_{tot} - E_1)_{(eo)} dE_1 \quad (6)$$

Similar equations hold for even-odd and odd-odd fissioning systems. In the reduced energy scale used in eqs. (3-6), the level densities of neighboring even-even, odd-A and odd-odd nuclei are very similar for positive reduced excitation energies (see Fig. 3). Therefore, the difference between the number of configurations $N_{Z_1=e}$ and $N_{Z_1=o}$ is mainly given by the different limits of the integrals. For even-even fissioning nuclei, eqs. (3) and (4) show that $N^{ee}_{Z_1=e} > N^{ee}_{Z_1=o}$ for every charge split $Z_1$ and $Z_2$ and the even-odd effect $\delta_p$ will always be positive. In contrast, for odd-even fissioning nuclei $N^{oe}_{Z_1=e} \approx N^{oe}_{Z_1=o}$ close to symmetry because $\Delta_1 \approx \Delta_2$. $N^{oe}_{Z_1=e}$ starts to be larger than $N^{oe}_{Z_1=o}$ as we move to more asymmetric splits because $\Delta_1 > \Delta_2$.

In practice one cannot always find experimental level densities representative of even-even, odd-A or odd-odd nuclei for each value of $Z_1$ and $Z_2$ covered by the fission yields. However, given the similarity between the experimental level densities of neighboring nuclei, we have replaced in eqs. (3-6) the representative level densities by the level densities $\rho_i$ of the two fission fragments considered, namely $A_1$, $Z_1$ and $A_2$, $Z_2$. The level densities $\rho_i$ are obtained using a composite formula with a constant-temperature description at low excitation energies and a shifted Fermi-gas description above. In order to include the effects of pairing correlations on the nuclear level density in a realistic way, the recommended energy shift of the Fermi-gas part of the Gilbert-Cameron composite formula [25] is increased by 2 MeV [26]. Consequently, in our formula the transition from the constant-temperature to the Fermi-gas regime occurs at energies of about 8-9 MeV which are higher than the transition energies of the broadly used Gilbert-Cameron formula. The logarithmic slope of the Fermi-gas level density becomes more gradual with increasing excitation energy, and the relative statistical weight of configurations with a fully-paired light fragment is less important than in the constant-temperature regime. Therefore, the transition from the constant-temperature to the Fermi-gas regime that may occur when $E_{tot}$ increases will lead to a considerable decrease of $δ_p$. This explains why $δ_p$ decreases with increasing mass of the fissioning nucleus. In our description, we assume that the level densities follow the constant-temperature behavior down to $-Δ$ for odd-$A$ and down to $-2Δ$ for even-even nuclei. As shown in Fig. 3, this is rather reasonable for most of the fission fragments. The reason is the presence of collective levels and the energy resolution of the measurements. The latter leads to a kind of averaging that can be justified by the fluctuations of the energy exchange [4].

The results of our calculation are compared with experimental data in Fig. 2. The increase of $δ_p$ with asymmetry and, except for $^{230}$Th, the decrease of $δ_p$ with increasing mass of the fissioning nucleus are fairly well reproduced. For $^{236}$U, the data point that is closest to symmetry is appreciably higher than the calculation. This effect may be associated to the influence of the $Z=50$ shell in the complementary fragment, which is known to enhance the yield of tin isotopes. Our calculation is also in good agreement with the data for the odd-Z fissioning nucleus $^{243}$Am. We obtain best agreement with fission data (including a wide variety of fission observables and fisioning nuclei [27]) when we consider that 40% of the potential energy difference from saddle to scission [28] is dissipated. The dissipated energy varies from about 4 MeV for $^{230}$Th to 11 MeV for $^{250}$Cf. When $E_{tot}$ is small $δ_p$ varies very rapidly with $E_{tot}$. Therefore, $^{230}$Th is particularly sensitive to the uncertainties on the dissipated energy. The disagreement found for $^{230}$Th may be caused by the neglect of fluctuations in the dissipated energy. In fact, for a great part of the fission events the available energy may be so low that they reach the scission point in a completely paired configuration due to the threshold character of the first quasi-particle excitation.

The formation of a fully-paired light fragment takes time. This time is the sum of the time needed for the light fragment to transfer its energy to the heavier one and the time to transfer one or two unpaired nucleons through the neck. If this time is longer than the saddle-to-scission time, our model will over predict the magnitude of the even-odd effect. Therefore, the agreement between the experimental data and our calculation indicates that the time to form a fully-paired light fragment is shorter than the saddle-to-scission time. This is an interesting finding that puts constraints on the time scales of the shape evolution and of excitation-energy transport in large-amplitude collective motions under the influence of pairing correlations, which represents still a considerable challenge. In principle, the energy dissipated at neck rupture adds up to the excitation energies of the fragments in fixed proportions, given by the geometry of the necking configuration. However, we expect that this additional excitation energy does not influence the even-odd effect as predicted by our model because there is not enough time at scission to distribute this excitation energy and subsequently transfer nucleons between the fragments according to statistical equilibrium.

In conclusion, we have shown that the general dependence of the even-odd effect in fission-fragment yields with the mass of the fissioning system and with the mass asymmetry of the fragments can be explained by the influence of statistical mechanics on the fission process, which favors the production of even-even light fragments with no quasi-particle excitations. This implies that excitation energy and unpaired nucleons are predominantly transferred to the heavy fragment. Thus, the even-odd effect in fission-fragment yields represents a second strong evidence for the importance of the entropy-driven energy-sorting mechanism in low-energy fission, which is fully consistent with the excitation-energy dependence of the prompt-fission neutron yields [3].

This work was supported by the European Commission within the Sixth Framework Programme through EFNUDAT (project no. 036434) and within the Seventh Framework Programme through Fission-2010-ERINDA (project no.269499).

---------------------------------------------------------------

[1] M. Guttormsen et al. , Phys. Rev. C **68**, 034311 (2003)
[2] M. Guttormsen et al., submitted to Phys. Rev. C
[3] K.-H. Schmidt, B. Jurado, Phys. Rev. Lett. **104**, 212501 (2010)
[4] K.-H. Schmidt, B. Jurado, Phys. Rev. C **83**, 014607 (2011)
[5] K.-H. Schmidt, B. Jurado, Phys. Rev. C **83**, 061601 (R) (2011)
[6] K.-H. Schmidt et al., Nucl. Phys. A **665**, 221 (2000)
[7] V. M. Strutinsky, Nucl. Phys. A **122**, 1 (1968)
[8] B. L. Tracy et al., Phys. Rev. C **5**, 222 (1972)
[9] M. Caamano, F. Rejmund, K.-H. Schmidt, J. Phys. G: Nucl. Part. Phys. **38**, 035101 (2011)
[10] E. Wigner, Trans. Faraday Sot. **34**, part 1, 29 (1938)
[11] U. Mosel and H. W. Schmitt, Nucl. Phys. A **165**, 73 (1971)
[12] M. Mirea, Phys. Lett. B **680** (2009) 316
[13] H. J. Krappe, Int. J. Mod. Phys. E **16**, 396 (2007)
[14] W. D. Myers and W. J. Swiatecki, Nucl. Phys. A **612**, 249 (1997)
[15] D. H. Gross, Entropy **6**, 158 (2004) / Special Issue "Quantum Limits to the Second Law of Thermodynamics"
[16] A. C. Larsen et al., Phys. Rev. C **83**, 034315 (2011)
[17] A. Schiller et al., Phys. Rev. C **63**, 021306 (R) (2001)
[18] U. Agvaanluvsan et al., Phys. Rev. C **70**, 054611 (2004)
[19] A. Bürger et al., Phys. Rev. C **85**, 064328 (2012)
[20] A.C. Larsen et al., Phys. Rev. C **76**, 044303 (2007)
[21] N.U.H. Syed et al., Phys. Rev. C **80**, 044309 (2009)
[22] M. Guttormsen et al., Phys. Rev. C **83**, 014312 (2011)
[23] A.C. Larsen et al., Phys. Rev. C **73**, 064301 (2006)
[24] T. von Egidy, D. Bucurescu, Phys. Rev. C **72**, 044311 (2005)
[25] R. Capote et al., Nucl. Data Sheets **110**, 3107 (2009)

[26] K.-H. Schmidt, B. Jurado, Phys. Rev. C **86**, 044322 (2012)
[27] The GEneral Fission Model (GEF): www.cenbg.in2p3.fr/GEF
[28] M. Asghar and R. W. Hasse, J. Phys. Colloques **45**, C6-455 (1984)